\documentstyle[prl,aps,floats,epsfig]{revtex}
\newcommand{\be}{\begin{equation}}
\newcommand{\ee}{\end{equation}}
\newcommand{\bees}{\begin{eqnarray}}
\newcommand{\ees}{\end{eqnarray}}
\newcommand{\ra}{\rightarrow}

\newcommand{\ls}{\lambda_{\rm s}}
\newcommand{\dd}[2]{\nabla_{#1}\nabla^{#2}}
\newcommand{\pz}{\phi_0}
\newcommand{\vp}{\varphi}

\begin{document}

\par
\begingroup

\begin{flushright}
 IFUP-TH 10/97\\
  May 1997\\
  gr-qc/9705082
\end{flushright}

{\large\bf\centering\ignorespaces
Density Perturbations in String  Cosmology
\vskip2.5pt}
{\dimen0=-\prevdepth \advance\dimen0 by23pt
\nointerlineskip \rm\centering
\vrule height\dimen0 width0pt\relax\ignorespaces
Michele Maggiore$^{a,b}$ and Riccardo Sturani$^{b,c}$
\par}
{\small\it\centering\ignorespaces

(a) INFN, sezione di Pisa, Pisa, Italy\\
(b) Dipartimento di Fisica dell'Universit\`{a},
piazza Torricelli 2, I-56100 Pisa, Italy\\
(c) Scuola Normale Superiore, 
piazza dei Cavalieri 7,I-56125 Pisa, Italy.\par}

\par
\bgroup
\leftskip=0.10753\textwidth \rightskip\leftskip
\dimen0=-\prevdepth \advance\dimen0 by17.5pt \nointerlineskip
\small\vrule width 0pt height\dimen0 \relax

We discuss the generation and 
evolution of density perturbations during the large curvature 
phase of string cosmology. We find that perturbations in the scalar
components of the metric evolve with cosmic time as $\exp (\gamma
H_st)$ where  $H_s$ is the Hubble constant during the string phase and
$\gamma$ is a constant which is determined by an algebraic equation
involving all orders in the string coupling $\alpha '$.
The seed for the perturbations can be provided by massive string
modes.

\par\egroup

\thispagestyle{plain}
\endgroup

\section{Introduction}

A very important issue in any cosmological model is the generation of
density perturbations which evolved into the presently
observed large scale structures of the Universe.
At present, two main  mechanisms have been studied.
One is the amplification of zero point fluctuations at the transition
between two cosmological phases, as in the transition from an 
inflationary phase to a standard radiation dominated phase; and the
second possibility relies on topological defects 
produced at a GUT phase
transition (see e.g.~\cite{KT,HK}).

In the last few years, a cosmological model based on the low energy
effective action of string theory has been proposed by Gasperini and
Veneziano~\cite{Ven,GV,review}. The model addresses the kinematical problems of
standard cosmology thanks to a superinflationary 'pre-big-bang' 
phase (see however ref.~\cite{TW} for critical remarks) and to a
subsequent De-Sitter like inflationary stage, 
and has a number of interesting phenomenological 
consequences~\cite{review,BGGV,peak,BMU,AB,GGV}. 
The purpose of this paper is
to examine the issue of the generation and evolution of density
perturbations in this model. 

The analysis of refs.~\cite{GV,BGGMV}
indicates that the mechanism of amplification of vacuum fluctuations
is not effective in string cosmology at very large wavelengths
and it is probably unable to produce density fluctuations
at the level $\delta\rho /\rho\sim 10^{-5}$ as required for COBE
anisotropy and for galaxies formation. 
Of course, it is still
possible that density perturbations are generated by cosmic strings or
other topological defects at the GUT phase transition, or in general 
after the end of the high curvature  phase of the model.
In this paper, however, we explore the possibility 
that a genuinely 'stringy' effect in the high curvature regime is
responsible for the generation of density fluctuations.

The paper is organized as follows. In sect.~2 we briefly recall the main
features of the cosmological model. In sect.~3 we compute the evolution
of scalar perturbations during the high curvature regime in a
simplified setting. (The details of the full analysis are relegated in
the Appendix.) In sect.~4
we discuss a possible mechanism for the generation of the seed for
scalar perturbation. Sect.~5 contains the conclusions.

\section{A short summary of string cosmology}

In this section we present the main features of the model and we
establish the notations. For a more detailed presentation we refer the
reader to refs.~\cite{Ven,GV,review}. The model is obtained 
by considering the  Neveu-Schwarz bosonic sector
of string theory.
The low-energy effective action  is built from the
massless states of the string: the graviton $g_{\mu\nu}$, 
the dilaton $\phi$ and the antisymmetric field $B_{\mu\nu}$ and, in
the string frame,  it reads
\be\label{act}
S^{(0)}=-\frac{1}{2\ls^2}
\int d^4x\sqrt{-g}e^{-\phi}\left[ R+(\nabla\phi )^2-
\frac{1}{12}H^2\right]\, ,
\ee
where $H_{\mu\nu\rho}=\partial_{[\mu}B_{\nu\rho ]}$, $R$ is the
Ricci scalar\footnote{We use the  sign
conventions $\eta_{\mu\nu} = (+,-,-,-)$ and
$R^{\mu}_{\,\,\nu\rho\sigma}
=\partial_{\rho}\Gamma^{\mu}_{\nu\sigma}-\ldots\, $.}
and $\ls$ is the string length. This action
receives two kinds of corrections~\cite{review}. The analogue of
quantum field theory loop corrections, which are governed by
$e^{\phi}$, and genuinely stringy corrections which are controlled by 
the string constant $\alpha'$ (related to the string length $\ls$ and
to the string tension $\mu$ by $\sqrt{2\alpha '}=\ls =1/\sqrt{\pi\mu}$).

In the pre-big-bang scenario the evolution starts from
the string perturbative vacuum at low curvature and $\phi\ra -\infty$,
where both kind of corrections are neglegible and physics is well
described by the effective action~(\ref{act}), and it follows a phase of
superinflationary evolution, which ends in a singularity at a finite
value of cosmic time. 
However, when the curvature reaches a value of order
one in string units, $\alpha '$ corrections to eq.~(\ref{act}),
e.g. terms $\alpha ' R_{\mu\nu\rho\sigma}^2$, become important. As
shown in~\cite{GMV}, at first order in $\alpha' $ they do indeed stop
the growth of the curvature: the effective 
action at order $\alpha'$ receives a correction which depends on the
Gauss-Bonnet term $R_{\rm GB}^2$ \cite{Zwi} 
and after a field redefinition at order
$\alpha '$ it can be written in the form
\be\label{act1}
S^{(1)}=-\frac{1}{2\ls^2}
\int d^4x\sqrt{-g}e^{-\phi}\left[ R+(\nabla\phi )^2
-\frac{\alpha '}{4}\left( R_{\rm GB}^2-(\nabla\phi )^4\right)
\right]\, ,
\ee
where we have reabsorbed
into $\alpha'$ a factor $k=1,1/2$ for bosonic and heterotic string
respectively and we have  assumed that the torsion background
$B_{\mu\nu}$ is trivial.

The action~(\ref{act1}) has a solution of the equations of
motion  of the form $H={\rm const.}=H_s$, 
$\dot{\phi}={\rm const.}=c$, 
where $H$ is the Hubble constant, $H_s,c$ are 
numerical constants of order $1/\ls$, and the dot is the derivative
with respect to cosmic time. This solution acts as a late time
attractor of the lowest order pre-big-bang solution obtained from
eq.~(\ref{act}). 
Of course, when terms $O(\alpha ')$  are
crucial, also terms $O(\alpha '^2)$, etc.  become crucial.
Therefore we are in a truly stringy regime. The general
structure of the equations of motions at all orders in $\alpha '$ is
discussed in ref.~\cite{GMV}, where it is found that a solution of the
form $H={\rm const.}=H_s$, $\dot{\phi}={\rm const.}=c$, persists at all
orders in $\alpha '$ if a set of two  {\em algebraic}
equations in the two unknown $c,H_s$,  involving
all orders in $\alpha '$, has real solutions. In the following, we
assume that this is indeed the case.

Finally, this 'string phase' should be matched to the standard radiation
dominated phase. The problems connected with
implementing this transition
('graceful exit') have been investigated recently 
by various authors~\cite{graceful}.

\section{Evolution of density perturbations during the string phase}

Metric perturbations around a FRW background can be classified
according to their properties under spatial rotations. Scalar
perturbations, which couple to density fluctuations, can be 
written in the longitudinal gauge in terms of two scalar functions
$\psi_1,\psi_2$ as~\cite{Muk}
\be \label{pert}
ds^2=a^2(\eta )\left\{ (1+2\psi_1 )d\eta^2 -(1-2\psi_2 )d{\bf x}^2
\right\}\, ,
\ee
where $a$ is the scale factor, $\eta$ is conformal time, and we are
considering a spatially flat 4-dimensional FRW background. 
In this section, to illustrate our results, we limit ourselves to
fluctuations with $\psi_1= -\psi_2$, i.e. to metrics which can be 
written in terms of a  single function $h$ as
\be
g_{\mu\nu}(\eta,{\bf x})=\left[
1+h(\eta,{\bf x})\right]\bar{g}_{\mu\nu}(\eta )
\ee
where $\bar{g}_{\mu\nu}=a^2(\eta )\eta_{\mu\nu}$ 
is the background metric. The fact that these fluctuations are simply
fluctuations in the scale factor allows a substantial simplifications
of the relevant equations and allows us to present equations which are
not too lengthy.  As it will be clear below, 
we are only interested in general properties of the solutions, and we
will show in the Appendix that these properties still hold if we
consider the general case with $\psi_1$ and $\psi_2$ independent. 

We also expand the dilaton as $\phi (\eta,{\bf x})=\phi_0(\eta )
+\varphi (\eta,{\bf x})$. 
We make these substitutions into eq.~(\ref{act1}) and expand to
second order in $h,\varphi$. The condition that the linear terms vanish
give the dynamical equations of motion for the background, 
\be\label{1}
\bar R-(\nabla\pz)^2+2\nabla^2\pz-\alpha'\left[\frac{1}{4}
\left(\bar R^2_{GB}+3(\nabla\pz)^4\right)-(\nabla\pz)^2
\nabla^2\pz-2\nabla^\mu\pz\nabla^\nu\pz\nabla_\mu\nabla_\nu
\phi_0\right]=0
\ee
\be\label{2}
\bar{R}-2(\nabla \phi_0 )^2+3\nabla^2\phi_0
-\alpha '\bar{G}^{\mu\nu}(\nabla_{\mu}\phi_0\nabla_{\nu}\phi_0
-\nabla_{\mu}\nabla_{\nu}\phi_0)=0\, .
\ee
(The overbar denotes quantities constructed with $\bar{g}_{\mu\nu}$,
and $\bar{G}^{\mu\nu}$ is the Einstein tensor. Indices are raised and
lowered with the background metric, and covariant derivatives 
$\nabla_{\mu}$ are with respect to the background metric.
The first equation is obtained with
a variation with respect to $\varphi$ and the second with respect to
$h$.)  Furthermore,
the variation with respect to the lapse function gives 
a constraint on the initial values.\\
For our homogeneous background, 
$\bar g_{\mu\nu}=a^2(\eta)\eta_{\mu\nu},\phi_0=\phi_0(\eta)$
the dynamical equations of motion, eqs.~(\ref{1},\ref{2}), become
\bees\label{1b}
2\ddot\pz-\dot\pz^2+6\dot\pz\frac{\dot a}{a}-6\frac
{\dot a^2}{a^2}-6\frac{\ddot a}{a}-\alpha '\left\{6\frac{\ddot a}{a}
\frac{\dot a^2}{a^2}+\frac{3}{4}\dot\pz^4-3\dot\pz^3\frac
{\dot a}{a}-3\dot\pz^2\ddot\pz\right\}=0,\\
\label{2b}
3\ddot\pz-2\dot\pz^2+9\dot\pz\frac{\dot a}{a}-6\frac{\dot
a^2}{a^2}
-6\frac{\ddot a}{a}-\alpha'\left\{-6\dot\pz\frac{\dot a}{a}\frac{\ddot a}{a}
+3\dot\pz^2\frac{\dot a^2}{a^2}-3\ddot\pz\frac{\dot a^2}{a^2}-3\dot\pz
\frac{\dot a^3}{a^3}\right\}=0,
\ees
while the constraint equation is
\be
\label{lapse}
-6\frac{\dot a^2}{a^2}-\dot\pz^2+6\dot\pz\frac{\dot a}{a}+\alpha'
\left\{6\dot\pz\frac{\dot a^3}{a^3}-\frac{3}{4}\dot\pz^4\right\}=0\, .
\ee
Eqs.~(\ref{1b},\ref{2b},\ref{lapse}) have the solution
$H=H_s,\dot{\phi}=c$ which, in terms of conformal time, reads
\be\label{back}
a(\eta )=-\frac{1}{H_s\eta},\hspace*{10mm}
\phi_0(\eta)=-\frac{c}{H_s}\log (-H_s\eta)\hspace*{15mm} (\eta <0)\, .
\ee
Notice that the ansatz~(\ref{back}) 
reduces eqs.~(\ref{1b},\ref{2b},\ref{lapse})
to 3 equations in 2 unknown $c,H_s$; remarkably, on this ansatz the 3
equations are not independent and admit a common solution. This goes
through at all orders in $\alpha '$ due to general properties of the
beta functions of the underlying sigma model~\cite{GMV}.

Expanding the action at second order in $\varphi ,h$
and then varying it, or linearizing eqs.~(\ref{1},\ref{2}),
we get the equations of motion for the fluctuations, which read
\bees\label{3}
& &2\nabla^{\mu}\phi_0\nabla_{\mu}\varphi-2\nabla^2\varphi +3\nabla^2h
-2\nabla^{\mu}\phi_0\nabla_{\mu}h
-h\left(\bar{R}-(\nabla\phi_0)^2+2\nabla^2\phi_0\right) +\nonumber\\
& &-\alpha '\left[ 
-\bar{G}^{\mu\nu}\nabla_{\mu}\nabla_{\nu}h +
\nabla_{\mu}\varphi \left(
4\nabla_{\nu}\phi_0\nabla^{\mu}\nabla^{\nu}\phi_0
-3\nabla^{\mu}\phi_0(\nabla\phi_0)^2+2\nabla^{\mu}\phi_0\nabla^2
\phi_0\right)+ \right.\\
& & \left.
+\nabla_{\mu}\nabla_{\nu}\varphi \left(
2\nabla^{\mu}\phi_0\nabla^{\nu}\phi_0
+\bar{g}^{\mu\nu}(\nabla\phi_0)^2\right)\right]=0\, ,\nonumber
\ees
and
\bees\label{4}
& &-4\nabla^{\mu}\phi_0\nabla_{\mu}\varphi+3\nabla^2\varphi -3\nabla^2h
+3\nabla^{\mu}\phi_0\nabla_{\mu}h+\nonumber\\
& &-\alpha '  
 \left[ 
(\nabla_{\mu}\nabla_{\nu}h )
  \left(\bar{R}^{\mu\nu}
   +\nabla^{\mu}\nabla^{\nu}\phi_0-\nabla^{\mu}\phi_0\nabla^{\nu}\phi_0
   -\bar{g}^{\mu\nu}(\nabla^2\phi_0-(\nabla\phi_0)^2) \right)+ 
\right. \\
& &+(\nabla_{\mu}h )
  (\frac{1}{2}\nabla^{\mu}\bar{R}-\bar{R}^{\mu\nu}
   \nabla_{\nu}\phi_0)
+h\bar{G}^{\mu\nu} 
  (\nabla_{\mu}\nabla_{\nu}\phi_0-\nabla_{\mu}\phi_0\nabla_{\nu}\phi_0)
+2\nabla_{\mu}\varphi\bar{G}^{\mu\nu}\nabla_{\nu}\phi_0
-\bar{G}^{\mu\nu}\nabla_{\mu}\nabla_{\nu}\varphi
{\large \left. \right] }
=0\, .\nonumber
\ees
(In the above expressions we have not yet used the specific form of the
background.) The linearization of the constraint equation is of course
satisfied with the initial conditions $h =\varphi =0$.

Although formally correct, eqs.~(\ref{3},\ref{4}) suffer from a serious
ambiguity. Consider for instance the 
coefficient of the term  $\alpha ' h$ in
eq.~(\ref{4}), i.e.
$\bar{G}^{\mu\nu}  
(\nabla_{\mu}\nabla_{\nu}\phi_0-\nabla_{\mu}\phi_0\nabla_{\nu}\phi_0)$.
We are free to add to it a term which vanishes because of
the equations of motions of the background, 
e.g. an arbitrary constant times the
left-hand side of eq.~(\ref{1}). The latter, however, is
made of two separately non zero terms, one of order zero in $\alpha '$
and one of order $\alpha'$. 
In a formal expansion in powers of $\alpha '$
of the coefficient of $h$ in eq.~(\ref{4}), part of the term that we
have added would contribute to order $\alpha '$ and part to order
$\alpha '^2$; at the order at which we are working, 
this second part would be discarded. Thus, the coefficient of $h$ in
eq.~(\ref{4}), at order $\alpha '$, could be equally well written as 
\be
-\alpha'\left[ \bar{G}^{\mu\nu}
(\nabla_{\mu}\nabla_{\nu}\phi_0-\nabla_{\mu}\phi_0\nabla_{\nu}\phi_0)
+ c_1 (\bar{R}-(\nabla \phi_0 )^2+2\nabla^2\phi_0)\right]\, ,
\ee
with $c_1$ an arbitrary constant.
The same holds for all other operators in
eqs.~(\ref{3},\ref{4}); for instance the coefficent of the term
$\nabla_{\mu}\nabla_{\nu}h$ can be modified adding a term proportional
to $\bar{g}^{\mu\nu}$ times the left-hand side of
eq.~(\ref{1}). Therefore, a truncation of the coefficients at a finite
order in $\alpha '$ gives ambiguous results.

One might try to argue that there is at least
a 'natural' prescription for the coefficients at any finite order
which consists of taking eqs.~(\ref{3},\ref{4}) as they are, without
adding terms proportional to the equations of motion. However, even this
is  not true. Consider for instance the term 
$(\nabla^2\phi_0-(\nabla\phi_0)^2)$ which appears at order $\alpha '$
in eq.~(\ref{4}). If we insert the explicit expression for $\phi_0$,
eq.~(\ref{back}), this is equal to $3cH_s-c^2$. However, if we combine
eqs.~(\ref{1}) and (\ref{2}) we find that 
$(\nabla^2\phi_0-(\nabla\phi_0)^2)$ is $O(\alpha ')$ and, since this
term appears in eq.~(\ref{4}) already multiplied by $\alpha' $, it
gives a contribution $O(\alpha '^2)$. Clearly, there is no natural
choice between these two ways of treating this term.

What we learn, therefore, is that a truncation at any finite order in
$\alpha '$ gives completely arbitrary results. Only the full answer,
including all orders in $\alpha '$, is meaningful. This is a
consequence of the fact that the solution around which we are
expanding emerges only after $\alpha '$ corrections are taken into
account, and it is not a solution of  the lowest order effective action,
eq.~(\ref{act}). So, the background solution mixes terms of different
order in $\alpha '$. 

We can however still use eqs.~(\ref{3},\ref{4}) in order to learn
something on the structure of the equations at all orders. Using the
explicit form of the background, eq.~(\ref{back}), we find that
eqs.~(\ref{3},\ref{4}) can be recast in the general form
\be\label{eq}
\eta^2\left( a_1^{(i)}\frac{d^2h}{d\eta^2}-
a_2^{(i)}\vec{\nabla}^2h\right) +a_3^{(i)}h+
a_4^{(i)}\eta \frac{dh}{d\eta}+
\eta^2\left( b_1^{(i)}\frac{d^2\varphi}{d\eta^2} -
b_2^{(i)}\vec{\nabla}^2\varphi \right)
+b_3^{(i)}\eta\frac{d\varphi}{d\eta}=0\, ,
\ee
where  $\vec{\nabla}$ is the flat space
spatial gradient and the index $i=1,2$ enumerates the two equations
obtained from eqs.~(\ref{3}) and (\ref{4}), respectively.
The coefficients $a_1^{(i)}$, etc. are independent of $\eta$ and
are formally given as a power expansion
in $\alpha '$. From eqs.~(\ref{3},\ref{4}) we formally get for instance
$a_1^{(2)}=-3H_s^2+\alpha '(3H_s^4-cH_s^3)$, etc., but as explained before,
a truncation at finite order in $\alpha '$ is actually meaningless. 
Note that this would be true even if, for some numerical accident,
$H_s,c$ were much smaller than one in units $\alpha '=1$.

Let us introduce the 
modes $h_k(\eta )$ performing the Fourier transform of
$h(\eta, {\bf x})$ with respect  to ${\bf x}$, and similarly 
for $\varphi$. Consider in eq.~(\ref{eq}) the term
$a_1{d^2h}/d\eta^2-a_2\vec{\nabla}^2h$, which can be rewritten
as 
$a_2[ (1/c_s^2)d^2h_k/d\eta^2+k^2h_k]$
where $c_s=(a_2/a_1)^{1/2}$ represents, physically the speed of the
perturbation and, on general grounds, $c_s<1$. For wavelengths with
$\eta k c_s<1$ the gradient term $k^2h_k$
is small compared to $(1/c_s^2)d^2h_k/d\eta^2\sim h_k/(c_s^2\eta^2)$.
Of course, this is just the condition that the physical wavelength of
the perturbation is greater than the Jeans length. 
On the other hand, if $c_s<1$, there is a window of values of the
wavelength for which at the same time the perturbation is still
within the horizon, $k\eta >1$.  The latter condition
is important because for
super-horizon-sized modes it becomes relevant the fact that the
scalar perturbations $\psi_1,\psi_2$ are not exactly
gauge-invariant. Indeed, $\psi_1,\psi_2$ are  the Bardeen
variables written in the longitudinal gauge~\cite{Bar,Muk}, 
and are gauge-invariant only at first order for
small coordinate transformations. The effect of non gauge invariance
becomes important for physical wavelength larger than the horizon, 
$\lambda_{\rm phys}>H^{-1}$, where $\lambda_{\rm phys}=a(\eta
)\lambda$. In a DeSitter background this means $k\eta <1$ where
$k=1/\lambda$ is the comoving wavenumber. For $k\eta <1$ a
growing  perturbation can be a gauge artefact with no physical
meaning. 

If instead $1<k\eta <1/c_s$
there is a range of wavelengths for which all
gradient terms can be neglected but the perturbations are still within
the horizon; then  we see that
equation~(\ref{eq}) for $h_k,\varphi_k$ depends
on $\eta$ and $d/d\eta$ only through
the combination $\eta d/d\eta$. This was not   obvious 
{\em a priori}. On
dimensional grounds, $\eta$ could also have appeared in combinations
as $\eta H_s,\eta c$, and therefore the coefficients $a_1^{(i)}$,
etc., could have been functions of $\eta$, while instead they only
depend on the constants $H_s,c,\alpha '$.
This result can  be traced to the specific
form of the De Sitter background and of the dilaton,
eq.~(\ref{back}). To understand this point, observe 
for instance that in this background, using conformal time,
the  Ricci tensor is given by 
$\bar{R}^{\mu\nu}=-3H_s^4\eta^2\eta^{\mu\nu}$, while 
$\bar{g}^{\mu\nu}=H_s^2\eta^2\eta^{\mu\nu}$, 
$\partial^{\mu}\phi_0=\bar{g}^{\mu\nu}\partial_{\nu}\phi_0=
-cH_s\eta\delta_0^{\mu}$ and 
similarly $\nabla^{\mu}\nabla^{\nu}\phi_0\sim\eta^2$. So, in a term like
$\nabla_{\mu}\nabla_{\nu}h$ in eqs.~(\ref{3},\ref{4}), 
independently of whether we use
$\bar{R}^{\mu\nu},\bar{g}^{\mu\nu}$ 
or derivatives with respect to $\phi_0$ to
saturate the indices $\mu ,\nu$, we get a
factor of $\eta$ for each index to be saturated. One can easily check
that the argument goes through for all operators one can construct,
and therefore it holds at all orders in $\alpha '$. 
For instance, at higher orders we will have terms with more than two
derivatives,
e.g.
$\nabla_{\mu}\nabla_{\nu}\nabla_{\rho}\nabla_{\sigma}h$. Whatever
operator we use to saturate the four indices, we get a term
$\sim\eta^4d^4h/d\eta^4$. 

The existence of a range of values of $k$ for which the perturbation
is within the horizon, $k\eta >1$, but still the gradient term can be
neglected, is a general properties due basically to the fact that the
speed of the scalar perturbation ('speed of sound') is smaller than
the speed of light, and we expect that it  goes through at all
orders in $\alpha '$, when higher order spatial and temporal
derivatives come into play. 
If this is the case then, for such wavelengths,
the full equations governing the perturbations, at all orders in
$\alpha '$, have the general form
\be\label{FG}
F_i(\eta\frac{d\,}{d\eta})h_k(\eta )+
G_i(\eta\frac{d\,}{d\eta})\varphi_k(\eta )=0\, ,
\ee
where $i=1,2$ enumerate the equation and $F_i,G_i$ are unknown
functions. The non-trivial content of eq.~(\ref{FG}) is that they
depend on $\eta$ only through the combination $\eta d/d\eta$.
Looking for solutions of the form $h_k\sim\eta^{\gamma}$,
$\varphi_k\sim\eta^{\beta}$ reduces  eqs.~(\ref{FG}) to a set of two
{\em algebraic } equations for the two unknown $\beta ,\gamma$:
\be
F_1(\gamma )+G_1(\beta)=0,\hspace{5mm}F_2(\gamma )+G_2(\beta)=0\, .
\ee
The situation is particularly interesting if the above equations have
a real and negative solution for $\gamma$. In fact, since in DeSitter
space the relation between conformal and cosmic time is 
$-H_s\eta  = \exp (-H_s t)$, we get an exponentially growing
mode
\be\label{growing}
h_k\sim \exp\{|\gamma |H_st\}\, .
\ee
The maximum amplification which can be obtained is  
related to the duration of the string phase. Denoting with $\eta_s$
the value of conformal time at which the string phase begins and with
$\eta_1$ the value at which it ends,$ (\eta_s,\eta_1 <0,
|\eta_s|>|\eta_1|)$  the 
maximum amplification factor is $ (\eta_s/\eta_1)^{|\gamma |}$. The
parameter $ (\eta_s/\eta_1)$ (equal to $f_1/f_s$ in the notation of
ref.~\cite{BMU}) is a free parameter of the model, and can be
very large. In fact, the most interesting 
phenomenological situation from the point of
view of the production of relic gravitational waves is realized when
this number is at least of order $10^8$~\cite{BGGV,BMU}.

We see therefore that string cosmology has a 
possible built-in mechanism 
which allows to amplify in a substantial way very tiny initial
inhomogeneities. 

Of course, the above result only holds in the linear regime. However,
this is sufficient for our purpose, which is to find a mechanism which
generates density fluctuations at the level
$(\delta\rho/\rho )_k\sim h_k\sim 10^{-5}$ at wavelength
of the order of the present Hubble radius or at the scale of galaxies
formation.

\section{The seed for perturbations}

In the previous section we found that, 
depending on the form of some unknown algebraic
functions $F_i,G_i$, which involve a full (all-orders in $\alpha '$)
computation, we can have a very large enhancement of initial density
fluctuations. The next step is to identify a possible candidate
mechanism which generates a seed which is then amplified. 

In the dilaton dominated regime where the action~(\ref{act}) gives a
good description of physics, there is no mechanism which can generate
inhomogeneities. In fact, as recently shown by Veneziano~\cite{Ven2},
even starting with generic inhomogenehous initial conditions the 
model evolves into a highly
homogeneous spacetime.  Therefore, we have to
look again into the string phase in order to find a seed for density
perturbations. 

Until now we have limited our attention to the massless modes of the
string. However,  massive string modes are associated with $\alpha '$
corrections to the action~\cite{Tse,BFLP}, and must  be included
for consistency once we include corrections of the type 
$\alpha ' R_{\mu\nu\rho\sigma}^2$. In fact the complete 
effective action of the string has the general form~\cite{BFLP}
\be\label{str}
S=\int d^2\sigma\sum_{n=0}^{\infty}(\alpha ')^n
\sum_{i_n=1}^{N_n}{\cal
O}_{i_n}^{(n)}(\partial X) B_{i_n}^{(n)}(X)\, ,
\ee
where the string coordinates have been rescaled, $X^{\mu}\ra
\sqrt{\alpha '}X^{\mu}$, $B_{i_n}$ are the background field at a given
massive level (rescaled so that they are dimensionless)
and ${\cal O}_{i_n}^{(n)}$ the corresponding operators; $N_n$ is  the
number of irreducible operators for given $n$.
The power of $\alpha '$ is uniquely connected with the mass level. 
At least working order by order in the mass level~\cite{BFLP}, 
the condition of
quantum Weyl invariance of the action~(\ref{str}) induces $\alpha '$
corrections to the equations of motion derived from eq.~(\ref{act}).
Since we have seen that $\alpha '$ corrections have a crucial effect
in the string phase, it is clear that in this phase
it is not appropriate to restrict the attention to the
graviton-dilaton-antisimmetric tensor field sector. 

Of course, a description of the full stringy regime is very difficult
to obtain. However, a number of general properties are well understood
from the study of strings at very large temperature and 
densities~\cite{finiteT}. The main results are as follows.
At low temperatures and densities the canonical ensemble provides a
good description of a gas of strings. As $T$ approaches the Hagedorn
temperature $T_H$, the canonical ensemble breaks down and one has to resort
to the more fundamental microcanonical ensemble. As  $T\ra T_H^-$  the
canonical energy density tends to a finite value $\rho_c$
(in dimensions $d\ge 4$).
The most intriguing result is that if we increase the energy density
beyond this critical value, all the  excess energy goes into a single
highly excited string, rather than raising the temperature of the
string ensemble.

In the high curvature phase of string cosmology the parameter which
controls the value of the curvature is $H_s$. If $H_s$ exceeds a
critical value $H_c$, 
the energy density of the gravitational field exceeds
the critical energy density $\rho_c$ and the excess energy goes into a
very excited string.
More precisely, we should expect the formation of
one very excited string in each
horizon volume, since regions separated by more than a horizon
distance behave independently.\footnote{Here we are using the common
abuse of language of calling $H^{-1}$ the horizon distance. Actually,
in string cosmology the particle horizon is infinite~\cite{GV}, but still
$H^{-1}$ gives the scale of the distance between points which formally
recede from each other at the speed of light.}
This means that, if $H_s>H_c$, the favored thermodynamical
configuration is highly inhomogeneous, and 
the massive string modes can act as the seed for density perturbations.

To understand more in detail the effect of the highly excited strings,
let us denote by $\Theta_{\mu\nu}$ their energy momentum tensor.
It acts effectively as  an external source, and the
equation of motion of the perturbations obtained with a variation with
respect to $g_{\mu\nu}$ or, in our case, with respect to $h$ (which
corresponds to $i=2$ in eq.~(\ref{FG}))
becomes, at all orders in $\alpha '$,
\be\label{FG2}
F_2(\eta\frac{d\,}{d\eta})h_k(\eta )+
G_2(\eta\frac{d\,}{d\eta})\varphi_k(\eta )=\ls^2e^{\phi}\Theta\, ,
\ee
where $\Theta =\bar{g}^{\mu\nu}\Theta_{\mu\nu}$, while the equation with
$i=1$ remains homogeneous. 

As already mentioned,
eqs.~(\ref{1},\ref{2}), and therefore their linearization,
eqs.~(\ref{3},\ref{4}), do not exaust the  content of the classical
theory. Another equation is obtained varying the action with respect
to the lapse function $N$, which is then set equal to one with a
choice of gauge. This equation, of course, is a constraint on the
initial data, which is conserved by the dynamical equations of
motions. The linearization of the constraint equations  gives as
the initial conditions for the perturbations $h=\varphi =0$ (and
similarly all the time derivatives of $h,\varphi$ can be set to zero up
to one order less than the higher order derivative appearing in the
dynamical equations). Since the energy momentum tensor is coupled to
the lapse function, the constraint equation in the presence of massive
string modes  has the general form $C(h,\varphi)=\Theta$ where $C$ is
a function of $h,\varphi$ and of their derivatives which vanishes when
all its arguments are zero.

Therefore, we see that the 'external source' $\Theta$ has a two-fold
effect. First, it forces $h$ to have a non-zero initial value, through
the constraint equation. And second, the evolution of the perturbation
will be the sum of a particular solution of the inhomogeneous
equation~(\ref{FG2}) and of the general solution of the associated
homogenehous equation,  which includes (if $\gamma <0$) the
exponentially growing mode~(\ref{growing}). These two effects, which
are quite general, correspond to what have been termed 
as 'initial compensation' and 'subsequent compensation' in
ref.~\cite{VS}. 

It is quite difficult to estimate the form of the spectrum of density
perturbations, since it depends on the details of the string
phase, and we leave the problem open for further work.
Even more difficult is to estimate  the numerical value of 
$(\delta\rho/\rho )_k$, since it involves in an essential way the
numerical values of $\eta_s,\eta_1,\gamma$ on which we have no
clue. 

\section{Conclusions}

The problem of the generation and evolution of density perturbations
in string cosmology appears to be a very difficult one, since it
fully involves the details of the high curvature stringy regime, for
which we do not have an adequate theoretical understanding.

Still, a number of general considerations appear to be possible. 
It turns out that in a number of situations some dynamical questions
concerning the string phase and which therefore
 in principle are very complicated, can 
actually be reduced to the issue of
the existence of a solution to a certain {\em algebraic} equation
involving all orders in $\alpha '$.

An example of this phenomenon
has been found in ref.~\cite{GMV}, where the issue of the existence of
a solution of the equations of motion which forbids the lowest order
pre-big bang solutions from running into a singularity
 has been shown to reduce, to all orders 
in $\alpha '$,  to the existence of a real solution
of an algebraic equation.
In this paper we have found a similar characterization for the
existence of exponentially growing scalar perturbations during the
string phase. The basic mechanism is that, for wavelengths larger than
the Jeans length, in a DeSitter background with a
dilaton growing linearly with cosmic time, the equations of motion of
the perturbation have solutions of the form $\eta^{\gamma}$ and, in
DeSitter space, $\eta\sim \exp (-H_st)$; this results, for $\gamma <0$,
in an exponentially growing mode.

{}From the physical point of view,  both the exponential
growth of the perturbation, and the generation of seeds from highly
excited strings appear under rather general circumstances
and it is therefore possible that string cosmology has its own,
genuinely stringy mechanism for generating 
large scale density perturbations.

As already stressed in~\cite{GMV}, an all order solution should
correspond to an exact conformal field theory 
(see~\cite{KK} for work along these lines) and our algebraic
conditions could have a natural counterpart in the conformal field
theory approach. 

\vspace{5mm}

We thank Maurizio Gasperini and Gabriele Veneziano for useful comments.

\appendix

\section{}

In this Appendix we examine the equations governing the evolution of
the perturbations in the general case with $\psi_1$ and $\psi_2$
independent. Technically, we lose the advantage of having a metric
conformal to Minkowski metric, and as a consequence most of our
equations become very lengthy. However we are only
interested in the general properties of these equations, and in
particular in the fact that they only depend on $\eta d/d\eta$ once
spatial gradients have been neglected.

The equations of motion for the background are of course unchanged. We
now have three dynamical
equations of motion for the fluctuations, which can be
obtained by
expanding the action at second order and performing the
variations with respect to $\varphi ,\psi_1,\psi_2$. 
The variation w.r.t. $\vp$ gives 
\bees\label{a1}
& &2\nabla_\mu\vp
\nabla^\mu\pz-2\nabla^2\vp+2\nabla_\mu(\psi_1+\psi_2)\nabla^\mu\pz+
4\nabla_0(\psi_1+\psi_2)\nabla^0\pz-4\nabla_\mu\psi_1\nabla^\mu\pz
-6\nabla^2\psi_2+2\dd kk(\psi_1+\psi_2)+\nonumber\\
& & +(\psi_1+3\psi_2)\bar R_0^0-
(\psi_1-\psi_2)\bar R^i_i+(\psi_2-\psi_1)(2\nabla^2\pz-\nabla_\mu
\nabla^\mu\pz)+2(\psi_1+\psi_2)(2\dd 00\pz-\nabla_0\pz\nabla^0\pz)+
\nonumber\\
& & -\frac{\alpha'}{4}\{
8\dd \mu\nu\bar G^\mu_\nu-R^{2\ (1)}_{GB}+
3\bar R^2_{GB}(\psi_1+\psi_2)+4(\nabla\pz)^2(\nabla^2\vp
-3\nabla_\nu\pz\nabla^\nu\vp) 
+16\dd \mu\nu\pz\nabla^\mu\pz\nabla_\nu\vp+\\
& & +8\nabla^2
\pz\nabla_\mu\vp\nabla^\mu\pz+8\dd \mu\nu\vp\nabla^\mu\pz\nabla_\nu\pz
+(\psi_1+\psi_2)\left[-36\dd 00\pz\nabla_0\pz\nabla^0\pz
+9(\nabla\pz)^4+\right.\nonumber\\
& &\left.\left.-4\dd ii\pz\nabla_0\pz\nabla^0\pz\right]
-12\nabla_0(\psi_1+\psi_2)(\nabla^0\pz\nabla_0\pz\nabla^0\pz)
\right\}=0\nonumber,
\ees
where repetead latin indices like $i,j,k$\ldots \ mean summation
over the three spatial coordinates and the index $0$ denotes the time
coordinate. 
The variation w.r.t. $\psi_1$ gives 
\bees
& & 2\dd ii
\vp-2\nabla_\mu\vp\nabla^\mu\pz+3(\psi_1+\psi_2)\bar R_0^0-(\psi_1
+\psi_2)\bar R_i^i+\nonumber\\
& &+(\psi_1+\psi_2)\left[(\nabla\pz)^2
+2\nabla_0\pz\nabla^0\pz-2\dd kk\pz\right]+4\dd kk \psi_2-
2\nabla_\mu\pz\nabla^\mu\psi_2-4\nabla_0\pz\nabla^0\psi_2+\nonumber\\
& &-\frac{\alpha'}{4}\{3\vp\bar R^2_{GB}+9\bar R^2_{GB}(\psi_2
-\psi_1)+40\,\bar G^\mu_\nu\dd\mu\nu\psi_2+32\nabla^2\psi_2\bar R-
8\nabla_0\psi_2\nabla^0\pz\bar R
+8(\psi_1+\psi_2)\bar R^{0i}_{0j}(\bar R^{0j}_{0i}+\bar R^j_i)+
\nonumber\\
& &
+4(\psi_1+\psi_2)\left(\bar R^{ki}_{lj}\right)^2-8(\psi_1+\psi_2)\left(
\bar R^{0i}_{kj}\right)^2+8(\psi_1+\psi_2)\left(\bar R^{ki}_{0j}\right)^2
-8\left(\dd ij (\psi_1+\psi_2)\right)\bar R^{0i}_{0j}
-8(\psi_1+\psi_2)
\left(\bar R_0^0\right)^2+\nonumber\\
& &-16(\psi_1+\psi_2)\left(\bar R^i_j\right)^2
-32(\psi_1+\psi_2)\left(\bar R^0_i\right)^2+8\left(\dd ij (\psi_1
+\psi_2)\right)\bar R^i_j+4(\psi_1+\psi_2)\bar R\bar R^i_i+\nonumber\\
& &+4\left(\dd ii(\psi_1+\psi_2)\right)(\bar R^0_0-\bar R^k_k)
+\psi_1\left[8\bar R^{0i}_{0j}(\bar R_{0i}^{0j}+
\bar R_i^j)+4(\bar R^{ki}_{lj})^2-8(\bar R^{0i}_{kj})^2+8(\bar R^{ki}
_{0j})^2+\right.\nonumber\\ \label{a2}
& & \left.+8\left(\bar R_{0i}^{0j}-\bar R_i^j\right)\dd ji\pz
-8(\bar R_0^0)^2-16(\bar R_i^j)^2-32(\bar R^i_0)^2+4\bar R\bar R^i_i
-4\left(\dd kk\pz\right)(\bar R_0^0-\bar R_i^i)\right]+\\
& &+8(5\psi_1-\psi_2)
\left(\bar R_{0i}^{0j}\right)^2+\bar R_{0i}^{0j}\left[16\dd ji\psi_1-
8\dd ji\psi_2
+16\dd ji\pz(2\psi_2-\psi_1)-32\psi_1\bar R_j^i\right]+\nonumber\\
& &+32(4\psi_1+
\psi_2)\left(\bar R^{ki}_{lj}\right)^2
+\bar R_0^0\left[-24\nabla^2\psi_2+40\dd kk\psi_2-16\dd kk\psi_1
+36\nabla_\mu\psi_2\nabla^\mu\pz-8\bar R_0^0(5\psi_1-\psi_2)+\right.
\nonumber\\
& & \left.
+16\psi_1\dd kk \pz-8\bar R^i_i(4\psi_1+\psi_2)-24\psi_2\dd kk\pz
\right]
+\bar R^i_j\left[-56\dd ij\psi_2-16\dd ij\psi_1+16(\psi_1-2\psi_2)\dd ij\pz
\right]+\nonumber\\
& &+\bar R_i^i\left[28\nabla^2\psi_2+16\nabla_0\psi_2\nabla^0\pz
-24\psi_1\bar R_i^i+8\psi_2\dd jj\pz\right]
+\bar R\left[-48\nabla^2\psi_2+8\dd ii\psi_1-8(\psi_1-\psi_2)\dd ii\pz
+24\psi_1\bar R\right]+\nonumber\\
& &-8\dd ij\dd ji\psi_1+8\left(\dd ij(\psi_1-\psi_2)\right)
\left(\dd ji\pz\right)-16\left(\dd jj \psi_2\right)
\left(\dd ii\pz\right)+24\left(\nabla^2\psi_2\right)
\left(\dd ii\pz\right)+\nonumber\\
& &-24\dd ii\dd 00\psi_2+24\bar R^\mu_\nu \dd \mu\nu\psi_2
-24\left(\nabla^\mu\psi_2\right)\bar R_\mu^\nu\nabla_\nu\pz+\nonumber\\
& &-3(\nabla_0\pz\nabla^0\pz)^2(5\psi_1+3\psi_2)+12(\nabla_0\vp
\nabla^0\pz)(\nabla_0\pz\nabla^0\pz)-3\vp(\nabla_0\pz\nabla^0\pz)^2
\}=0.\nonumber
\ees
To simplify the equations, in the terms between curly brackets we have
used the fact that $\phi_0$ does not depend on spatial coordinates,
and  that during the stringy phase every component of
Riemann and Ricci tensor with an equal number of upper and lower indices 
is costant in time as well as in space.

The equation that one would obtain by varying the second order 
action w.r.t. $\psi_2$ can  be obtained more simply  noting that
subtracting it from eq.~(\ref{a2}) 
one gets equation (\ref{4}).

Having all the equations, 
we can now directly check by inspection that the
property of homogeneity is indeed preserved. If we neglect all spatial
gradients, the  equations~(\ref{a1}), (\ref{a2})
and (\ref{4})  have the general form
\be\label{FG3}
A_i(\eta\frac{d}{d\eta})\psi_{1k}(\eta )+
B_i(\eta\frac{d}{d\eta})\psi_{2k}(\eta )+
C_i(\eta\frac{d}{d\eta})\varphi_{k}(\eta )=0,
\ee
where now $i=1,2,3$, and therefore we can look for solutions
$\psi_1\sim\eta^{\gamma_1},\psi_2\sim\eta^{\gamma_2},
\varphi\sim\eta^{\beta}$, which reduce eqs.~(\ref{FG3}) to a set of 3
algebraic equations for 3 unknowns $\gamma_1,\gamma_2,\beta$. The
general situation is therefore completely analogous to that discussed
in the text.

\end{document}